\documentclass[
reprint,              % preprint: one column; reprint: two columns
nofootinbib,          % No footnotes in bibliography
superscriptaddress,   % Use superscripts for affiliations
showpacs,             % Show PACS numbers
showkeys,              % Show keywords
longbibliography    % Show paper titles in bibliography 
]{revtex4-1}

\usepackage[utf8]{inputenc}
\usepackage{amsmath,amssymb,amsthm} % amsthm must be loaded *after* amsmath
\usepackage{bm,dsfont} % bbm package uses Type 3 fonts - avoid!
\usepackage{url}
\usepackage{graphicx}
\usepackage{multirow}
\usepackage{siunitx}
\usepackage{xcolor}
\usepackage{hyperref}
\hypersetup{linktocpage,colorlinks=true,urlcolor=blue,linkcolor=blue,citecolor=blue}

\usepackage{tikz}

\usetikzlibrary{
	babel,
	quotes,
	arrows.meta,
	shapes.symbols,
	shadows.blur,
}

\tikzset{
	> = LaTeX,
	pics/lisa/.style = {
		code = {
			\draw [densely dashed, thick] (0, 1) -- (210:1) -- (330:1) -- cycle;
			\foreach \th in {90, 210, 330} {
				\draw [thick, fill = white] (\th:1) circle [radius = 0.2];
			}
		}
	}
}

\definecolor{lime}{HTML}{A6CE39}
\DeclareRobustCommand{\orcidicon}{
	\begin{tikzpicture}
	\draw[lime, fill=lime] (0,0) 
	circle [radius=0.16] 
	node[white] {{\fontfamily{qag}\selectfont \tiny ID}};
	\draw[white, fill=white] (-0.0625,0.095) 
	circle [radius=0.007];
	\end{tikzpicture}
	\hspace{-2mm}
}

\foreach \x in {A, ..., Z}{%
\expandafter\xdef\csname orcid\x\endcsname{\noexpand\href{https://orcid.org/\csname orcidauthor\x\endcsname}{\noexpand\orcidicon}}
}

\theoremstyle{plain}

\theoremstyle{definition}

\newcommand{\udec}{Departamento de Fí­sica, Universidad de Concepción, Casilla 160-C, Concepción, Chile}

\newcommand{\unap}{ Instituto de Ciencias Exactas y Naturales (ICEN), Facultad de Ciencias, Universidad Arturo Prat, Iquique, Chile}
\newcommand{\asctp}{Universidad Austral de Chile (UACh), Campus Isla Teja, Ed. Emilio Pugin, Valdivia, Chile}

\newcommand{\cecs}{Centro de Estudios Científicos (CECs), Avenida Arturo Prat 514, Valdivia, Chile}

\begin{document}

\title{Mimetic Einstein-Cartan-Kibble-Sciama (ECKS) gravity}

%\affiliation{\unap}
%\affiliation{\icen}

\author{Fernando Izaurieta}
\email{fizaurie@udec.cl}
\affiliation{\udec}

\author{Perla Medina}
\email{perlamedina@udec.cl}
\affiliation{\udec}
\affiliation{\cecs}
\affiliation{\asctp}

\author{Nelson Merino}
\email{nemerino@unap.cl}
\affiliation{\unap}
%\affiliation{\icen}

\author{Patricio Salgado}
\email{pasalgad@unap.cl}
\affiliation{\unap}
%\affiliation{\icen}

\author{Omar Valdivia\orcidA{}}
\email{ovaldivi@unap.cl}
\affiliation{\unap}
%\affiliation{\icen}
\date{\today}

\bigskip

\begin{abstract}
In this paper, we formulate the Mimetic theory of gravity in first-order formalism for differential forms, i.e., the mimetic version of Einstein-Cartan-Kibble-Sciama (ECKS) gravity. We consider different possibilities on how torsion is affected by conformal transformations and discuss how this translates into the interpolation between two different conformal transformations of the spin connection, parameterized with a zero-form parameter $\lambda$. We prove that regardless of the type of transformation one chooses, in this setting torsion remains as a non-propagating field. We also discuss the conservation of the mimetic stress-energy tensor and show that the trace of the total stress-energy tensor is not null but depends on both, the value of $\lambda$ and spacetime torsion. 

\end{abstract}



\maketitle

%%%%%%%%%%%%%%%%%%%%%%%%%%%%%%%%%%%%%%%%%%%%%%%%%%%%%%%%%%%%%%%%%%%%%%%%%%%%%%%%

%\tableofcontents

%%%%%%%%%%%%%%%%%%%%%%%%%%%%%%%%%%%%%%%%%%%%%%%%%%%%%%%%%%%%%%%%%%%%%%%%%%%%%%%%

\section{Introduction}
General Relativity (GR) is a classical field theory describing the gravitational interaction through the Einstein field equations. Remarkably, it has proven success in a wide range of phenomena~\cite{Will:2014kxa} including black-holes as realistic astrophysical objects~\cite{Akiyama:2019cqa} and the existence of gravitational waves~\cite{TheLIGOScientific:2017qsa,Monitor:2017mdv,GBM:2017lvd}. Another significant development of GR lies in the context of cosmology, in which an extension of Einstein's field equations, by the inclusion of an early inflationary stage and a cold dark matter contribution, is in good agreement with observational data~\cite{Akrami:2018odb}. This fact is somehow dramatic since we know very little about the nature of dark matter. For this reason, the problem of identifying dark matter candidates attracts attention from both cosmology and particle physics. There are several dark matter candidates, as weakly interacting massive particles, sterile neutrinos, axions, cold massive halo objects, and primordial black holes~\cite{Gelmini:2015zpa,Bertone:2004pz,Young2016}.
 
Given the difficulties for the standard cosmology models to describe the nature of dark matter (See, for instance,~\cite{Bull:2015stt}), there has been a popular trend for considering modified gravity models~\cite{DeFelice:2010aj,Olmo:2011uz,Capozziello:2011et,Nojiri:2017ncd,BeltranJimenez:2017doy,Heisenberg:2018vsk}. However, observational constraints make it hard to create a consistent model of modified gravity still compatible with the principles of GR.

Another exciting direction is to consider GR beyond the limits of Riemannian geometry. The canonical model generalizing GR is the Einstein-Cartan-Kibble-Sciama (ECKS) gravity. This modification came firstly with the works of Elie Cartan in 1922, before the discovery of spin. However, Cartan's model did not bring much attention until the late 1950s, where Sciama and Kibble rediscovered Cartan's results~\cite{RevModPhys.48.393,Hehl:1994ue}. The main feature of ECSK gravity is that it accounts for spacetime torsion (See~\cite{Trautman:2006fp}).
In ECSK, quantum mechanical spin acts as the source of torsion\footnote{It is important to stress that with spin, we refer only to intrinsic quantum mechanical spin, and we should not confuse this with the angular momentum density. This natural confusion has already led to some mistakes in the literature, see Ref.~\cite{Hehl:2013qga}}, in the same way as energy is a source of curvature~\cite{Blagojevic:2013xpa,Arcos:2005ec}. For this reason, torsion could have been relevant at the extremely high fermion densities of the early Universe~\cite{Poplawski:2011jz,Unger:2018oqo,Kranas:2018jdc,Poplawski:2010kb,Ivanov:2016xjm,Razina:2010bj,Palle:2014goa,Poplawski:2012qy}. However, in standard ECSK, the torsion two-form does not propagate in a vacuum, and it interacts very weakly with Standard Model fermions (See Chap. 8.4 of Ref.~\cite{SupergravityVanProeyen} and Refs.~\cite{Puetzfeld:2014sja,Carroll:1994dq,Boos:2016cey}). Thus, torsion could be a component of dark matter~\cite{Tilquin:2011bu,Alexander:2019wne,Magueijo:2019vmk,Barker:2020gcp,Alexander:2020umk,Izaurieta:2020xpk,Magueijo:2019vmk}.

More recently, in~\cite{Chamseddine:2013kea,Chamseddine:2014vna}, Chamseddine and Mukhanov have considered a different approach for addressing dark matter with the Mimetic gravity theory. This model shows that the conformal degree of freedom of the gravitational field becomes dynamical even in the absence of matter. This extra degree of freedom corresponds to the mimetic field's energy density, which mimics the stress-energy tensor of extra pressureless dust without needing dark matter particles. Moreover, Ref.~\cite{Momeni:2015gka} discusses whether mimetic cosmology can create late-time acceleration and the inflationary stage of the universe. Besides, many authors have considered different aspects of mimetic gravity during the last years with exciting results. For instance, in black holes~\cite{Gorji:2019rlm,Chamseddine:2019fog,Sheykhi:2019gvk,Chamseddine:2019pux,Sheykhi:2018ffj,Nashed:2018urj,Nashed:2018qag,Nashed:2018aai,Chen:2017col,Oikonomou:2016fxb,Oikonomou:2015lgy}, black strings~\cite{Sheykhi:2020fqf}, brane-world scenario~\cite{Nozari:2019shm,Sadatian:2019qvn,Sui:2020aim,Guo:2018tpo,Zhong:2017uhn,Sadeghnezhad:2017hmr}, among others. For a more exhaustive survey, see~\cite{Malaeb:2020ecm,Casalino:2019tho,Nozari:2019esz,Shen:2019nyp,Chothe:2019fvs,Ganz:2019vre,Khalifeh:2019zfi,Bezerra:2019bgt,deCesare:2019pqj,Solomon:2019qgf,Malaeb:2019rdl,Zhong:2018fdq,deCesare:2018cts,Casalino:2018wnc,Ganz:2018vzg,Li:2018uwg,Gorji:2018okn,Paston:2018orc,Bodendorfer:2018ptp,Chamseddine:2018gqh,Chamseddine:2018qym,Babichev:2018afx,Zhong:2018tqn,Brahma:2018dwx,Casalino:2018tcd,Langlois:2018jdg,Odintsov:2018ggm,Golovnev:2018icm,Mirza:2017afs,Haghani:2017mhi,Dutta:2017fjw,Nojiri:2017ygt,Gorji:2017cai,Volkmer:2017ems,Paston:2017das,Arroja:2017msd,Vagnozzi:2017ilo,Baffou:2017pao,Sebastiani:2016ras,Cognola:2016gjy,Myrzakulov:2015kda,Deruelle:2014zza} and references therein.

In this paper, we pursue constructing a mimetic theory of gravity in first-order formalism for differential forms.  For a sufficiently general family of conformal transformations for the affine connection, the resulting theory is equivalent to \textquotedblleft mimicking\textquotedblright\ the ECSK model. For the construction, we assume that the vierbein one-form $e^{a}(x)$ and the spin connection one-form $\omega^{ab}(x)$ describe independent metric and affine properties of the spacetime.
We consider diverse ways of generalizing conformal transformations for a Riemann-Cartan geometry with independent vierbein and spin connection, following a similar approach to~\cite{Chakrabarty:2018ybk}. Remarkably, none of these generalizations generate a propagating torsion field. Therefore, in this setting, only a non-vanishing spin tensor can be a source of torsion. When using scalar fields, non-minimal couplings and higher derivatives terms seem to be the way of creating a propagating torsion (See, for instance,~\cite{Valdivia:2017sat,PhysRevD.100.124039}).
When considering generalized conformal invariance on Riemann-Cartan geometry, the stress-energy tensor's trace, which usually vanishes for conformally invariant theories of gravity, has a non-zero value depending on torsion and on a parameter which characterizing conformal transformations for the spin connection.

This paper organization is the following. In section~\ref{sec1}, we summarize the main aspects of mimetic gravity. In section~\ref{sec2}, we revisit ECSK gravity, and we give a brief description of Cartan's first order formalism for differential forms. In section~\ref{sec3}, we introduce some useful mathematical tools to describe conformal structures in Cartan's first-order formalism. In section~\ref{sec4}, we derive the equations of mimetic gravity in first-order formalism, show how these correspond to the mimetic ECKS model, and analyze the conservation law for the mimetic stress-energy tensor. Finally, in section~\ref{sec6}, we discuss the stress-energy tensor's trace and its dependency on torsion and conformal parameter $\lambda$. The paper concludes in \ref{sec8} with a summary and comments regarding some possible physical applications.

\section{Review of Mimetic gravity}\label{sec1}

Mimetic Gravity was first introduced by A. Chamseddine and V. Mukhanov as a
theory of gravity which naturally exhibits conformal symmetry as internal
degree of freedom~\cite{Chamseddine:2013kea}. Let $M_{4}$ be a four
dimensional spacetime and let us consider a physical metric $g_{\mu \nu}$, with Lorentz signature $(-,+,+,+)$,  depending on an auxiliary metric $\bar{g}%
_{\mu \nu}$ and a scalar field $\phi$, namely%
\begin{equation}
g_{\mu \nu}=-\bar{g}^{\alpha \beta}\partial_{\alpha}\phi \partial_{\beta}\phi
\bar{g}_{\mu \nu}\,.\label{m0}%
\end{equation}
The metric $g_{\mu \nu}$ is invariant with respect to conformal transformations
of the auxiliary metric $\bar{g}_{\mu \nu}$, i.e., it remains unchanged after rescaling
\begin{equation}
\bar{g}_{\mu \nu}\rightarrow \Omega^{2}\left(  x\right)  \bar{g}_{\mu \nu
}.\label{Eq_transf_conf_aux}%
\end{equation}
Additionally, it follows from (\ref{m0}) that
\begin{equation}
g^{\alpha \beta}\partial_{\alpha}\phi \partial_{\beta}\phi=-1\,.\label{m7}%
\end{equation}
The resultant new degree of freedom associated with the transformation
(\ref{m0}) represents the longitudinal mode of
gravity which is excited even in the absence of any matter field configurations.

The canonical action of GR is rewritten by considering the physical metric
$g_{\mu \nu}$ as function of the scalar field $\phi$ and the auxiliary metric
$\bar{g}_{\mu \nu}$%
\begin{equation}
S=\frac{1}{c}\!\int\!\mathrm{d}^{4}x\sqrt{-g\!\left(  \bar{g}_{\mu \nu},\phi \right)\!
}\left[  \frac{1}{\kappa_{4}}\!\left(  \frac{1}{2}R\left(  \bar{g}_{\mu \nu}%
,\phi \right)  -\Lambda \right)\!  +L_{m}\right]\!, \label{m1}%
\end{equation}
where $\kappa_{4}=\frac{8\pi G}{c^{4}}$, $R$ is the Ricci scalar constructed from $g_{\mu\nu}$ and $L_{m}$ stands for the matter
Lagrangian. The action (\ref{m1}) is invariant under conformal transformation because it only depends on $g_{\mu \nu}$ which is conformally invariant under
(\ref{Eq_transf_conf_aux}). The resulting dynamics can be directly obtained by starting from the variation of (\ref{m1}) with respect to the physical metric $g_{\mu\nu}$, then expressing $\delta g_{\mu\nu} $ in terms of $\delta\bar{g}_{\mu\nu}$ and $\delta{\phi}$ and assuming that the last two are independent. Thus,
\begin{align}
G^{\mu \nu}-\kappa_{4}\mathcal{T}^{\mu \nu}+\left(  G-\kappa_{4}\mathcal{T}\right)  g^{\lambda \mu
}g^{\sigma \nu}\partial_{\lambda}\phi \partial_{\sigma}\phi &  =0\,,\label{m5}\\
\nabla_{\mu}\left[  \left(  G-\kappa_{4}\mathcal{T}\right)  \partial^{\mu}\phi \right]
&  =0\,,\label{m6}%
\end{align}
where $G_{\mu \nu}=R_{\mu \nu}-\frac{1}{2}g_{\mu \nu}R+\Lambda g_{\mu \nu}$ is the
Einstein tensor, $\mathcal{T}_{\mu \nu}$ the energy momentum tensor, and $G$, $\mathcal{T}$ denote
their respective traces. The dynamics given in (\ref{m5}) and (\ref{m6}) departs from pure GR.
In Ref.~\cite{Golovnev:2013jxa}, an equivalent
formulation of Mimetic Gravity has been proposed where, instead of introducing
$\phi$ through the reparametrization (\ref{m1}), the physical metric
$g_{\mu \nu}$ is directly used together with a constrained scalar field,
enforcing (\ref{m7}) through a Lagrange multiplier.

Taking the trace in (\ref{m5}), direct calculation
shows%
\begin{equation}
\left(  G-\kappa_{4}\mathcal{T}\right)  \left(  1+g^{\alpha \beta}\partial_{\alpha}%
\phi \partial_{\beta}\phi \right)  =0\,.
\end{equation}
This last equation is automatically satisfied by the constraint (\ref{m7})
even for $G\neq \kappa_{4}\mathcal{T}$. From this point of view, even in absence of
matter, the gravitational field equations have non-trivial solutions for the
conformal mode. To understand this extra degree of freedom, let us rewrite
Eq.(\ref{m5}) as
\begin{equation}
G^{\mu \nu}=\kappa_{4}\left(  \mathcal{T}^{\mu \nu}+\mathcal{\bar{T}}^{\mu \nu}\right)\,,
\end{equation}
where
\begin{equation}
\mathcal{\bar{T}}^{\mu \nu}=\left(  \mathcal{T}-\frac{G}{\kappa_{4}}\right)  g^{\mu \alpha}%
g^{\nu \beta}\partial_{\alpha}\phi \partial_{\beta}\phi\,.
\end{equation}
Now compare this expression with the energy momentum tensor for a perfect
fluid%
\begin{equation}
\mathcal{T}^{\mu \nu}=\frac{1}{c^{2}}\left(  \varepsilon+p\right)  u^{\mu}u^{\nu}%
-pg^{\mu \nu}\,,\label{pft}%
\end{equation}
where $\varepsilon$ is the energy density, $p$ is the pressure and $u^{\mu}$
is the four-velocity which satisfies $\frac{1}{c^{2}}u^{\lambda}u_{\lambda
}=-1$. Setting $p=0$ and making the following identification
\begin{align}
&\varepsilon   =\left(\mathcal{T}-\frac{G}{\kappa_{4}}\right)\,,\\
&u^{\mu}   =c\,g^{\mu \alpha}\partial_{\alpha}\phi \,,
\end{align}
the energy momentum tensor (\ref{pft}) becomes equivalent to $\mathcal{\bar{T}}^{\mu \nu}$. Thus, the
extra degree of freedom mimics the potential motions of dust with energy
density $\left(\mathcal{T}-\frac{G}{\kappa_{4}}\right)$, and the scalar field plays the role of
a velocity potential. In absence of matter this energy density is proportional to
$G=4\Lambda-R$, which does not vanish for generic solutions. The
normalization condition for the four velocity $u^{\mu}$
and the conservation law for $\mathcal{\bar{T}}^{\mu \nu}$, are equivalent to
(\ref{m7}) and (\ref{m6}), respectively.

\section{ ECKS gravity and First order formalism}\label{sec2}

So far we have used Greek indices $\mu, \nu, \ldots$ to denote tensor
components in the coordinate basis. From now on, we use lower case Latin
indices $a, b, \ldots$ for tensors defined in Lorentz (orthonormal) basis. We
denote by $\Omega^{p}(M_{4})$ to the set of differential $p$-forms defined
over $M_{4}$.

At a particular point $P\in M_{4}$, the vierbein components  $e^{a}{}_{\mu}(x)$ are determined through the relation
\begin{equation}
g_{\mu \nu}=\eta_{ab}e^{a}{}_{\mu}e^{b}{}_{\nu}\,, \label{vier}%
\end{equation}
where $\eta_{ab}$ is the Minkowski metric. In terms of $e^{a}{}_{\mu}(x)$ we
define the vierbein $e^{a}=e^{a}{}_{\mu}(x)\mathrm{d}x^{\mu}$ as the set of
one-forms $\Omega^{1}(M_{4})\in T^{\ast}_{x}(M_{4})$. The vierbein contains
all the information on the metric.
%metric properties in such a way that one can shift from $g_{\mu \nu}$ to $e^{a}$ without any loss of generality.
 The one-form spin connection describes the affine properties of the geometry
$\omega^{ab}=\omega^{ab}{}_{\mu}(x)\mathrm{d}x^{\mu}$. The spin connection
$\omega^{ab}$ relates to $\Gamma_{~\mu \nu}^{\lambda}$ through the
vierbein postulate
\begin{equation}
\partial_{\mu}e^{a}{}_{\nu}+\omega^{a}{}_{b\mu}e^{b}{}_{\nu}-\Gamma^{\lambda
}{}_{\mu \nu}e^{a}{}_{\lambda}=0\,.
\end{equation}

The covariant derivative of the vierbein is defined as the two-form torsion
$T^{a}=\mathrm{D}e^{a}$ where
\begin{equation}
\mathrm{D}e^{a}=\mathrm{d}e^{a}+\omega^{a}{}_{b}\wedge e^{b}\,.%
\end{equation}
Unlike $\mathrm{d}^{2}=0$, higher order covariant derivatives of the vierbein
does not vanish. In fact, direct calculation shows $\mathrm{D}T^{a}=R^{a}%
{}_{b}\wedge e^{b}$ where
\begin{equation}
R^{ab}=\mathrm{d}\omega^{ab}+\omega^{a}{}_{c}\wedge \omega^{cb}\,,%
\end{equation}
is the Lorentz two-form curvature which transforms covariantly under local
Lorentz transformations.

The spin connection can also be decomposed in a torsion free part
$\mathring{\omega}^{ab}$ satisfying%
\begin{equation}
\mathrm{d}e^{a}+\mathring{\omega}_{~b}^{a}\wedge e^{b}=0\,,
\end{equation}
and a second rank anti-symmetric one-form $\kappa^{ab}$, usually called the
contorsion or contortion. An important observation is that $\mathring{\omega}^{ab}$ is
completely determined in terms of the vierbein. Therefore, all the
affine degree of freedom are encoded into the contorsion 
\begin{equation}
\kappa^{ab}%
=\omega^{ab}-\mathring{\omega}^{ab}\,,\label{fsp}
\end{equation}
 and consequently
$T^{a}=\kappa^{a}{}_{b}\wedge e^{b}$.
In terms of this splitting, the Lorentz curvature is given by
\begin{equation}
R^{ab}=\mathring{R}^{ab}+\mathring{\mathrm{D}}\kappa^{ab}+\kappa^{a}{}_{c}%
\wedge \kappa^{cb}\,,\label{lcc}%
\end{equation}
where $\mathring{R}^{ab}=\mathrm{d}\mathring{\omega}^{ab}+\mathring{\omega
}_{~c}^{a}\wedge \mathring{\omega}^{cb}$ is the Riemann curvature two-form and
$\mathring{\mathrm{D}}$ stands for the covariant derivative with respect to
the torsion free part of the connection $\mathring{\omega}^{ab}$.

There are many theories leading to non-vanishing torsion; see for instance Refs.~\cite{Carroll:1994dq,Alexander:2019wne,Magueijo:2019vmk,Barker:2020gcp,Valdivia:2017sat,Salgado:2014jka,Szabo:2014zua,Fierro:2014lka,Diaz:2012zza,Cid:2017wtf,Barrientos:2019awg,Alexander:2020umk,Toloza:2013wi,Pani:2009wy}, and for the most general case, Poincar\'{e} Gauge Theory, see~\cite{Hehl1980,Blagojevic:2003cg,Blagojevic:2012bc,Blagojevic:2013xpa,Obukhov:2018bmf,Obukhov:2020uan}). However, the closest to standard General Relativity is Einstein-Cartan-Kibble-Sciama (ECSK) gravity~\cite{Kib61,Sciama:1964wt,Hehl:1971qi,Hehl76,Shapiro:2001rz,Hammond:2002rm,Poplawski:2009fb}. In this framework, geometry also depends of the spin tensor of matter, but its physical effects would be noticeable only for densities much higher than nuclear density~\cite{Poplawski:2009su}.

Moreover, it has been shown that torsion
prevents
cosmological singularities~\cite{Kopczynski:1972fhu,KOPCZYNSKI197363,TRAUTMAN1973,Hehl:1974cn,DeSabbata1990,Gasperini:1986mv}, gives rise to a new universe from a collapse~\cite{Poplawski:2018ypb,Poplawski:2014dea,Poplawski:2011jz},
and introduce an effective ultraviolet cutoff in a quantum field theory for
fermions~\cite{Poplawski:2009su}. In differential form language, the ECKS four-form Lagrangian is given by
\begin{equation}
\mathcal{L}_{\mathrm{ECKS}}=\mathcal{L}_{\mathrm{G}}\left(  e,\omega \right)
+\mathcal{L}_{\mathrm{M}}\left(  e,\omega,\varphi \right)\,, \label{mac}%
\end{equation}
where
\begin{equation}
\mathcal{L}_{\mathrm{G}}=\frac{1}{4\kappa_{\mathrm{4}}}\epsilon_{abcd}\left(
R^{ab}-\frac{\Lambda}{3!}e^{a}\wedge e^{b}\right)  \wedge e^{c}\wedge e^{d}\,,%
\end{equation}
is the four-form Lagrangian for geometry and $\mathcal{L}_{\mathrm{M}}$ is the
four-form Lagrangian for matter fields. Up to boundary terms,
variation of the action functional $S=\frac{1}{c}\int_{M_{4}}\mathcal{L}%
_{\mathrm{ECKS}}$ reads
\begin{equation}
\delta S=\frac{1}{c}\int_{M_{4}}\frac{1}{\kappa_{\mathrm{4}}}\left(  \frac
{1}{2}\delta \omega^{ab}\wedge \mathcal{W}_{ab}+\mathcal{E}_{d}\wedge \delta
e^{d}\right)\,,
\end{equation}
with
\begin{align}
&  \mathcal{E}_{d}=\epsilon_{abcd}\left(  \frac{1}{2}R^{ab}-\frac{\Lambda}%
{3!}e^{a}\wedge e^{b}\right)  \wedge e^{c}-\kappa_{4}\ast \mathcal{T}_{d}%
\,,\label{mm10}\\
&  \mathcal{W}_{ab}=\epsilon_{abcd}T^{c}\wedge e^{d}-\kappa_{4}\ast \sigma
_{ab}\,.\label{mm11}%
\end{align}
Here, the relations
\begin{align}
\delta_{e}\mathcal{L}_{\mathrm{M}}^{\left(  4\right)  } &  =-\ast \mathcal{T}
_{d}\wedge \delta e^{d}\,,\\
\delta_{\omega}\mathcal{L}_{\mathrm{M}}^{\left(  4\right)  } &  =-\frac{1}%
{2}\delta \omega^{ab}\wedge \ast \,\sigma_{ab}\,.
\end{align}
with $\ast:\Omega^{p}\left(  M_{d}\right)  \rightarrow \Omega^{d-p}\left(
M_{d}\right)  $ denoting the Hodge dual operator
implicitly define the stress-energy one-form $\mathcal{T}^{a}=\mathcal{T}^{a}{}_{\mu}\mathrm{d}x^{\mu}$
and the spin tensor 1-form $\sigma^{ab}=\sigma^{ab}{}_{\mu}\mathrm{d}x^{\mu}$.

A usual practice when studying this system's phenomenology is to pack all the torsional terms (coming from the Lorentz curvature (\ref{lcc})) to create an extra stress-energy tensor in  (\ref{mm10}).

\section{Conformal Riemann-Cartan structure}\label{sec3}

To characterize conformal structures in differential forms language,
let us introduce an operator $\mathrm{I}_{a_{1}...a_{q}} :\Omega^{p}\left(
M_{d}\right)  \rightarrow \Omega^{p-q}\left(  M_{d}\right)  $ defined in four dimensional spacetime $\left(-,+,+,+\right)$
by~\cite{Valdivia:2017sat}
\begin{equation}
\mathrm{I}_{a_{1}...a_{q}} =-\left(  -1\right)  ^{p  \left(
p-q\right) }\ast e_{a_{1}}\wedge...\wedge e_{a_{q}}\wedge \ast \,.
\end{equation}
The case $q=1$ that gives
\begin{equation}
\mathrm{I}_{a}=-\ast e_{a}\wedge \ast \,,
\end{equation}
is of particular relevance because it satisfies useful properties.  It satisfies the Leibniz
rule for differential forms and, together with $\mathrm{D}$, the operator
$\mathrm{I}_{a}$ defines another important object $\mathcal{D}_{a}: \Omega^{p}
\left(  M_d \right)  \to \Omega^{p} \left(  M_d \right)  $ via the anti-commutator
\begin{equation}
\mathcal{D}_{a} = \left \{  \mathrm{I}_{a}, \mathrm{D} \right \}  =
\mathrm{I}_{a} \mathrm{D} + \mathrm{DI}_{a}\,, \label{Eq_Def_Da}%
\end{equation}
The operators $\mathcal{D}$, $\mathrm{I}_{a}$ and $\mathrm{D}$,
form an open superalgebra where the two-forms curvature and torsion play the role of
structure parameters~(See \cite{PhysRevD.100.124039}).

The conformal transformation
\begin{equation}
g_{\mu\nu}=\exp\left(  2\sigma\right)  \bar{g}_{\mu\nu}\,,\label{conf_T}%
\end{equation}
that relates the spacetime and the auxiliary manifolds supposes implicitly
that a local mapping $\gamma:{M_{4}\rightarrow\bar{M}_{4}}$ has been chosen in
such a way that the same coordinates $x^{\mu}$ can be used for $P\in$
$M_4$ and $\bar{P}=\gamma\left(  P\right)  \in\bar{M}_4$. This
means that a coordinate transformation $x^{\prime\mu}=x^{\prime\mu}\left(
x^{\nu}\right)  $ in $M_{4}$ induces the same transformation in
$\bar{M}_{4}$ and thus, tensors or forms defined on these manifolds
transforms with the same Jacobian matrices. This fact allows us to find the
relation between the vielbeins associated with these metrics, which by
definition satisfy%
\begin{align}
g_{\mu\nu} &  =e_{\mu}^{a}e_{\nu}^{a}\eta_{ab}\,,\\
\bar{g}_{\mu\nu} &  =\bar{e}_{\mu}^{a}\bar{e}_{\nu}^{b}\eta_{ab}\,.\
\end{align}
Indeed, mixing these expressions together with (\ref{conf_T}), it is direct to see that
\begin{equation}
e^{a}=\exp \left(  \sigma \right)  \bar{e}^{a}\,.\label{v0}%
\end{equation}

From the vierbein $e^{a}(x)$, it is possible to define \textquotedblleft structure parameters" $\mathcal{C}_{ab}^{~~c}(x)$ satisfying a generalized Maurer-Cartan equation.
\begin{equation}
\mathrm{d}e^{c}=-\frac{1}{2}\mathcal{C}_{ab}{}^{c}e^{a}\wedge e^{b}\,.\label{mc0}%
\end{equation}
This parameter allow us to solve the torsion-free part of the spin connection
\begin{equation}
\mathring{\omega}_{ab}=\frac{1}{2}\left(  \mathcal{C}_{abc}+\mathcal{C}%
_{cba}-\mathcal{C}_{cab}\right)  e^{c}\,.\label{mc1}%
\end{equation}
Using eqs.(\ref{v0}), (\ref{mc0}) and (\ref{mc1}), we find
\begin{equation}
\mathring{\omega}_{ab}=\overline{\mathring{\omega}}_{ab}+\bar{e}_{a}\xi
_{b}-\xi_{a}\bar{e}_{b}\,,\label{mmc1}%
\end{equation}
where
\begin{equation}
\xi^{a}=\bar{\mathrm{I}}^{a}\mathrm{d}\sigma\,,
\end{equation}
and
\begin{equation}
\mathrm{\bar{I}}_{a}=-\bar{\ast}\left(  \bar{e}_{a}\wedge \bar{\ast}\right. 
\end{equation}
with the bar in the Hodge dual denoting $\bar{e}^{a}$-vierbein dependence.
In this way, Eq.(\ref{mmc1}) characterizes the conformal transformation associated to
the torsion free part of the spin connection. 

Notice that we have no information yet on the conformal transformation of the contorsion $\kappa^{ab}$.
This is due to the fact that in the context of Riemann-Cartan geometry, $e^{a}$
and $\kappa^{ab}$ are completely independent degrees of freedom. Therefore,
there are multiple possible choices on how $\kappa^{ab}$ should transform
under a Weyl dilatation. An important family of choices can be parameterized as%
\begin{align}
\bar{e}^{a} &  \rightarrow e^{a}=\exp \left(  \sigma \right)  \bar{e}%
^{a}\,,\label{Eq_Gen_Conf_e}\\
\bar{\kappa}_{ab} &  \rightarrow \kappa_{ab}=\bar{\kappa}_{ab}+\left(
\lambda-1\right)  \theta_{ab}\,,\label{Eq_Gen_Conf_k}\\
\bar{\omega}_{ab} &  \rightarrow \omega_{ab}=\bar{\omega}_{ab}+\lambda
\theta_{ab}\,,\label{Eq_Gen_Conf_w}%
\end{align}
where $\lambda$ is a parameter $0\leq \lambda \leq1$ and $\theta^{ab}%
=-\theta^{ba}$ corresponds to the 1-form
\begin{equation}
\theta^{ab}=\bar{e}^{a}\xi^{b}-\xi^{a}\bar{e}^{b}\,.\label{teta}
\end{equation}
The case $\lambda=1$ implies,%
\begin{align}
\kappa_{ab} &  =\bar{\kappa}_{ab}\,,\\
\omega_{ab} &  =\bar{\omega}_{ab}+\theta_{ab}\,,
\end{align}
which is the \textquotedblleft canonical
case\textquotedblright: the full spin connection changes as the torsionless
case, and the contorsion is left untouched by the dilatation. The most
\textquotedblleft exotic\textquotedblright \ case corresponds to $\lambda=0$%
\begin{align}
\kappa_{ab} &  =\bar{\kappa}_{ab}-\theta_{ab}\,,\\
\omega_{ab} &  =\bar{\omega}_{ab}\,,
\end{align}
where the spin connection is left untouched by the dilatation and the
contorsion absorbs the transformation.

It is clear that the torsionless condition is preserved only for the
$\lambda=1$ case. In fact, the Lorentz curvature and torsion change under the
generalized Weyl dilatation (\ref{Eq_Gen_Conf_e}-\ref{Eq_Gen_Conf_w}) as%
\begin{align}
&\bar{T}_{a}   \rightarrow T_{a}=\exp \left(  \sigma \right)  \bar{T}%
_{a}+\left(  \lambda-1\right)  \bar{e}_{a}\wedge \mathrm{d}\exp \left(
\sigma \right)  ,\\
&\bar{R}^{ab}   \rightarrow R^{ab}=\bar{R}^{ab}+\lambda \mathrm{\bar{D}}%
\theta^{ab}+\lambda^{2}\theta^{a}{}_{c}\wedge \theta^{cb}.
\end{align}

\section{Mimetic ECKS gravity}\label{sec4}

Mimetic transformations are a particular choice of Weyl dilatations. Let us
consider the auxiliary vierbein $\bar{e}^{a}$ and spin connection $\bar
{\omega}^{ab}$ 1-forms, and a scalar field $\phi \left(  x\right)  $. In terms
%of $\bar{e}^{a}$ we define the operator $\mathrm{\bar{I}}_{a}:\Omega
%^{p}\left(  M^{\left(  4\right)  }\right)  \rightarrow \Omega^{p-1}\left(
%M^{\left(  4\right)  }\right)  $ by%
%\begin{equation}
%\mathrm{\bar{I}}_{a}=-\bar{\ast}\left(  \bar{e}_{a}\wedge %\bar{\ast}\right.  .
%\end{equation}
%where the bar in the Hodge dual denotes $\bar{e}^{a}$-vierbein dependence. In
of operators $\bar{\mathrm{I}}_{a}$ and $\phi$, let us define a zero-form Lorentz vector%
\begin{equation}
\bar{Z}_{a}=\mathrm{\bar{I}}_{a}\mathrm{d}\phi\,, \label{m10}%
\end{equation}
and the scalar%
\begin{equation}
\bar{Z}^{2}=-\eta_{ab}\bar{Z}^{a}\bar{Z}^{b}\,.
\end{equation}
The generalized mimetic vierbein $e^{a}$, contorsion $\kappa^{ab}$, and spin
connection $\omega^{ab}$ are defined by
\begin{align}
\bar{e}^{a}  & \rightarrow e^{a}=\bar{Z}\bar{e}^{a}\,,\label{vct}\\
\bar{\kappa}_{ab}  & \rightarrow \kappa_{ab}=\bar{\kappa}_{ab}+\left(
\lambda-1\right)  \theta_{ab}\,,\\
\bar{\omega}_{ab}  & \rightarrow \omega_{ab}=\bar{\omega}_{ab}+\lambda
\theta_{ab}\,,\label{Eq_w_mimetic}%
\end{align}
where the one-form $\theta^{ab}$ is given in (\ref{teta}) with
\begin{equation}
\xi^{a}=\frac{1}{\bar{Z}}\bar{\mathrm{I}}^{a}\mathrm{d}\bar{Z}\,.
\end{equation}
Notice that $\mathrm{I}_{a}$ and $\bar{\mathrm{I}}^{a}$ operator relate each other by
%\[
%\mathrm{I}_{a}:\Omega^{p}\left(  M^{\left(  4\right)  %}\right)  \rightarrow
%\Omega^{p-1}\left(  M^{\left(  4\right)  }\right)  ,
%\]
%given by%
\begin{equation}
\mathrm{I}_{a}  =\frac{1}{\bar{Z}}\bar{\mathrm{I}}_{a}\,,
\end{equation}
and consequently%
\begin{equation}
Z_{a}=\frac{1}{\bar{Z}}\bar{Z}_{a}\,,%
\end{equation}
so the constraint (\ref{m7}) reads
\begin{equation}
Z^{2}=-\eta_{ab}Z^{a}Z^{b}=1\,.\label{cme}
\end{equation}
\subsection{Mimetic field equations}\label{sec5}

To construct the mimetic version of ECSK theory, let us consider the
Lagrangian (\ref{mac}), with the vierbein $e^{a}$ and $\omega^{ab}$ in terms of
the auxiliary variables $\bar{e}^{a}$, $\bar{\omega}^{ab}$ and $\phi$ as in Eqs.~(\ref{vct})-(\ref{Eq_w_mimetic}). A priori, it would seem that different
choices of $\lambda$ would lead us to different dynamics. In particular, the
canonical and exotic choices $\lambda=1$ and $\lambda=0$ seem to lead to
completely different theories. However, nothing is further from truth. The
dynamics of the generalized mimetic theory is the same regardless of the
choice of $\lambda$.
Since%
\begin{align}
e^{a} &  =\bar{Z}\bar{e}^{a},\\
\omega_{ab} &  =\bar{\omega}_{ab}+\lambda \left(  \bar{e}_{a}\xi_{b}-\xi
_{a}\bar{e}_{b}\right)  ,
\end{align}
we have that the functional variations of the vierbein and the spin connection
are given by%
\begin{align}
\delta_{\bar{\omega}}e^{d} &  =0\,,\\
\delta_{\bar{e}}e^{d} &  =\delta_{\bar{e}}\bar{Z}\bar{e}^{d}+\bar{Z}\delta
\bar{e}^{d}\,,\label{Eq_de_e}\\
\delta_{\phi}e^{d} &  =\delta_{\phi}\bar{Z}\bar{e}^{d}\,,\label{Eq_d_phi_e}%
\end{align}
and%
\begin{align}
\delta_{\bar{\omega}}\omega_{ab}  & =\delta \bar{\omega}_{ab}\,,\label{Eq_dw_w}\\
\delta_{\bar{e}}\omega_{ab}  & =\lambda \left(  \delta \bar{e}_{a}\xi_{b}%
-\xi_{a}\delta \bar{e}_{b}\right)  +\lambda \left(  \bar{e}_{a}\delta_{\bar{e}%
}\xi_{b}-\delta_{\bar{e}}\xi_{a}\bar{e}_{b}\right)\,,\\
\delta_{\phi}\omega_{ab}  & =\lambda \left(  \bar{e}_{a}\delta_{\phi}\xi_{b}-\delta_{\phi}\xi_{a}\bar{e}_{b}\right)\,.
\end{align}
Notice that we need special care when performing the functional variation $\delta
\bar{Z}^{a}$. In fact, from definition (\ref{m10}), it is clear that $\bar
{Z}=\bar{Z}\left(  \bar{e},\partial \phi \right)  $. Therefore, we have to consider
independent variations of $\bar{Z}^{a}$ with respect to both, the vierbein
$\bar{e}^{a}(x)$ and the scalar field $\phi(x)$.
Since
\begin{equation}
\delta \bar{Z}=-\frac{1}{\bar{Z}}\bar{Z}^{a}\delta \bar{Z}_{a}\,,
\end{equation}
it is possible to prove that
\begin{align}
\delta_{\bar{e}}\bar{Z} &=\bar{Z}^{2}Z_{a}Z_{b}\mathrm{I}^{a}\left(  \delta \bar{e}^{b}\right)\,,
\label{Eq_de_Z}\\
\delta_{\phi}\bar{Z} 
&  =-\bar{Z}Z^{a}\mathrm{I}_{a}\mathrm{d}\delta \phi\,.\label{Eq_d_phi_Z}%
\end{align}
Replacing (\ref{Eq_de_Z})-(\ref{Eq_d_phi_Z}) into (\ref{Eq_de_e})%
-(\ref{Eq_d_phi_e}) we get the expressions%
\begin{align}
\delta_{\bar{\omega}}e^{d} &  =0\,,\\
\delta_{\bar{e}}e^{d} &  =\bar{Z}\left[  Z_{a}Z_{b}\mathrm{I}^{a}\left(
\delta \bar{e}^{b}\right)  e^{d}+\delta \bar{e}^{d}\right]  \,,\label{Eq_Def_de_e}%
\\
\delta_{\phi}e^{d} &  =-e^{d}Z^{a}\mathrm{I}_{a}\mathrm{d}\delta
\phi\,.\label{Eq_Def_d_phi_e}%
\end{align}

Up to boundary terms, variation of (\ref{mac}) reads
\begin{align}
&  \delta_{\bar{e}}\mathcal{L}_{\mathrm{ECSK}}^{\left(  4\right)  }=\frac
{1}{\kappa_{\mathrm{4}}}\left[  \frac{1}{2}\delta_{\bar{e}}\omega^{ab}%
\wedge \mathcal{W}_{ab}+\mathcal{E}_{d}\wedge \delta_{\bar{e}}e^{d}\right]
=0\,,\\
&  \delta_{\phi}\mathcal{L}_{\mathrm{ECSK}}^{\left(  4\right)  }=\frac
{1}{\kappa_{\mathrm{4}}}\left[  \frac{1}{2}\delta_{\phi}\omega^{ab}%
\wedge \mathcal{W}_{ab}+\mathcal{E}_{d}\wedge \delta_{\phi}e^{d}\right]  =0\,,\\
&  \delta_{\bar{\omega}}\mathcal{L}_{\mathrm{ECSK}}=\frac{1}{2}\frac{1}%
{\kappa_{\mathrm{4}}}\delta_{\bar{\omega}}\omega^{ab}\wedge \mathcal{W}%
_{ab}=0\,,\label{Eq_motion_mimetic_w}%
\end{align}
where the three-forms $\mathcal{E}_{a}$ and $\mathcal{W}_{ab}$ are given in
(\ref{mm10}) and (\ref{mm11}) respectively.
Since $\delta_{\bar{\omega}}\omega_{ab}=\delta \bar{\omega}_{ab}$,
Eq.(\ref{Eq_motion_mimetic_w}) implies $\delta \bar{\omega}^{ab}\wedge
\mathcal{W}_{ab}=0$ and consequently
\begin{equation}
\mathcal{W}_{ab}=0\,,
\end{equation}
just as in the standard ECSK model.
Inserting $\mathcal{W}_{ab}=0$ in the equations of motion, we are left with%
\begin{align}
&  \delta_{\bar{e}}\mathcal{L}_{\mathrm{G}}^{\left(  4\right)  }=\frac
{1}{\kappa_{\mathrm{4}}}\mathcal{E}_{d}\wedge \delta_{\bar{e}}e^{d}=0\,,\\
&  \delta_{\phi}\mathcal{L}_{\mathrm{G}}^{\left(  4\right)  }=\frac{1}%
{\kappa_{\mathrm{4}}}\mathcal{E}_{d}\wedge \delta_{\phi}e^{d}=0\,.
\end{align}
From here, using the expressions (\ref{Eq_Def_de_e}) and
(\ref{Eq_Def_d_phi_e}), and integrating by parts in $\mathrm{I}^{a}$ and
$\mathrm{d}$, we get the set of mimetic ECSK field equations%
\begin{align}
\mathcal{E}_{d}-Z_{a}Z_{d}\mathrm{I}^{a}\left(  \mathcal{E}_{m}\wedge
e^{m}\right)    & =0\,,\label{Eq_field_vierbein}\\
\mathrm{d}\left[  Z^{a}\mathrm{I}_{a}\left(  \mathcal{E}_{d}\wedge
e^{d}\right)  \right]    & =0\,,\label{Eq_field_scalar}\\
\mathcal{W}_{ab}  & =0\,.\label{Eq_field_w}%
\end{align}
It is remarkable that they do not depend on the choice of $\lambda$. For the
mimetic theory, all the choices of conformal transformations for the
contorsion lead to the same dynamics.

In order to study the equivalence of these equations written using tensors, it
is useful to consider Hodge duality between $p$-forms and $(d-p)$-forms. For a three-form $\mathcal{E}_{d}$ in four dimensions, we have
\begin{equation}
\mathcal{E}_{d}=\mathcal{E}_{md}\ast e^{m}.
\end{equation}
It is straightforward to prove that%
\begin{align}
\mathcal{E}_{q}\wedge e^{q}  & =-\mathcal{E}^{p}{}_{p}v_{\left(  4\right)
},\\
\mathrm{I}^{m}v_{\left(  4\right)  }  & =\ast e^{m},
\end{align}
where $v_{\left(  4\right)}$ denotes the volume form in four-dimensions.
Replacing these relations into the field equations (\ref{Eq_field_vierbein})%
-(\ref{Eq_field_w}) it is possible to write them as%
\begin{align}
\mathcal{E}_{d}-\ast \left(  Z_{d}\mathcal{E}^{p}{}_{p}\mathrm{d}\phi \right)
&  =0\,,\label{eom4}\\
-\mathrm{d}\ast \left[  \mathcal{E}^{p}{}_{p}\mathrm{d}\phi \right]   &
=0\,,\label{eom5}\\
\mathcal{W}_{ab} &  =0\,.\label{eom6}%
\end{align}
Remarkably, Eqs.~(\ref{eom4})-(\ref{eom5}) have the same form as Eqs.~(\ref{m5})-(\ref{m6}) but in terms of the full Lorentz curvature (\ref{lcc}) instead
of just the Riemannian piece. Note that Eq.~(\ref{eom6}) is the standard field
equation for torsion in terms of the spin tensor of matter.

\subsection{Conservation laws}\label{sec5'}

From (\ref{eom4}), it is clear that one can define a mimetic energy-momentum
one-form
\begin{equation}
\bar{\mathcal{T}}_{d}=\frac{1}{\kappa_{\mathrm{4}}}Z_{d}\mathcal{E}^{p}{}%
_{p}\mathrm{d}\phi\,.\label{memt}%
\end{equation}
The conservation law of $\bar{\mathcal{T}}_{d}$ in terms of the torsionless covariant derivative implies
\begin{align}
\mathrm{\mathring{D}}\ast \bar{\mathcal{T}}_{d} &    =\frac{1}{\kappa_{\mathrm{4}}}\mathrm{\mathring{D}}\left(  Z_{d}\ast \left[
\mathcal{E}^{p}{}_{p}\mathrm{d}\phi \right]  \right)  \nonumber\\
&  =\frac{1}{\kappa_{\mathrm{4}}}\left(  \mathrm{\mathring{D}}Z_{d}\wedge
\ast \left[  \mathcal{E}^{p}{}_{p}\mathrm{d}\phi \right]  +Z_{d}\wedge
\mathrm{\mathring{D}}\ast \left[  \mathcal{E}^{p}{}_{p}\mathrm{d}\phi \right]
\right) \nonumber \\
&  =\frac{1}{\kappa_{\mathrm{4}}}\left(  \mathcal{E}^{p}{}_{p}%
\mathrm{\mathring{D}}Z_{d}\wedge \ast \mathrm{d}\phi+Z_{d}\wedge \mathrm{d}%
\ast \left[  \mathcal{E}^{p}{}_{p}\mathrm{d}\phi \right]  \right)  .
\end{align}
Using (\ref{eom5}), we have
\begin{equation}
\mathrm{\mathring{D}}\ast \bar{\mathcal{T}}_{d}=\frac{1}{\kappa_{\mathrm{4}}%
}\mathcal{E}^{p}{}_{p}\mathrm{\mathring{D}}Z_{d}\wedge \ast \mathrm{d}\phi\,.
\end{equation}
Notice that
\begin{equation}
\mathrm{\mathring{D}}Z_{d}\wedge \ast \mathrm{d}\phi   =Z^{a}\mathcal{\mathring{D}}_{a}Z_{d}v_{\left(  4\right)  }\,,
%&  =\partial^{a}\phi \mathrm{\mathring{D}}_{a}\partial_{d}\phi v_{\left(  4\right)  }%
\end{equation}
and therefore, since $\mathcal{\mathring{D}}_{a}Z_{d}=\mathcal{\mathring{D}}_{d}Z_{a}$, one obtains
\begin{equation}
\mathrm{\mathring{D}}Z_{d}\wedge \ast \mathrm{d}\phi 
  =Z^{a}\mathrm{\mathring{D}}_{d}Z_{a}v_{\left(  4\right)  }
  =\frac{1}{2}\mathrm{I}_{d}\mathrm{d}(  Z^{a}Z_{a})  v_{\left(
4\right)  } =0\,,
\end{equation}
where we have used (\ref{cme}). Consequently,
\begin{equation}
\mathrm{\mathring{D}}\ast \bar{\mathcal{T}}_{d}=0\label{csl}\,,%
\end{equation}
which is the conservation law for the effective mimetic stress-energy tensor.

\section{The trace of the stress-energy tensor, torsion and $\lambda$}\label{sec6}

For the mimetic theory dynamics, the choice of the parameter $\lambda$ for the conformal transformations  (\ref{vct})-(\ref{Eq_w_mimetic}) seems to be
irrelevant. However, it does not mean that the parameter is meaningless.
In fact, it is related with the value of the trace of stress-energy tensor of
matter when its Lagrangian has conformal symmetry.

Let us consider a matter Lagrangian $\mathcal{L}_{\mathrm{M}}$ obeying conformal symmetry by itself%
\begin{equation}
\mathcal{L}_{\mathrm{M}}(e,\omega,\phi)  =\mathcal{L}%
_{\mathrm{M}}(  \Omega e^{a},\,\omega^{ab}+ \frac{\lambda}{\Omega}\left[e^{a},\mathrm{I}^{b}\right]\mathrm{d}\Omega,\frac{1}{\Omega^{\alpha}}\phi )\,.
\end{equation}
where $\left[e^{a},\mathrm{I}^{b}\right]=
e^{a}\mathrm{I}^{b}-e^{b}\mathrm{I}^{a}$. In the standard torsionless case, it would lead to a traceless on-shell
stress-energy tensor. This is no longer true in the current context of
non-vanishing torsion.
In fact, under an infinitesimal dilatation $\Omega=1+\varepsilon$, the field content of the matter Lagrangian changes according to%
\begin{align}
&\delta_{\varepsilon}e^{a}   =\varepsilon\, e^{a}\,,\label{ee1}\\
&\delta_{\varepsilon}\omega^{ab}   =\lambda \left[e^{a},\mathrm{I}^{a}\right]\mathrm{d}\varepsilon\,,\\
&\delta_{\varepsilon}\phi  =-\alpha\varepsilon \phi\,. \label{ee3}
\end{align}
Moreover, an arbitrary variation of $\mathcal{L}_{\mathrm{M}}(e,\omega,\phi)$ is given by%
\begin{align}
\delta \mathcal{L}_{\mathrm{M}}=&-\ast \mathcal{T}_{d}\wedge \delta e^{d}+\frac{1}{2}
\ast \sigma_{ab}\wedge \delta \omega^{ab}+\Phi\delta \phi\\&+\mathrm{d}%
\left(  \mathcal{B}_{a}^{\left(  2\right)  }\wedge \delta e^{a}+\frac{1}%
{2}\mathcal{B}_{ab}^{\left(  2\right)  }\wedge \delta \omega^{ab}+\mathcal{B}%
^{\left(  3\right)  }\delta \phi \right) \nonumber  ,
\end{align}
where $\Phi$ denotes the field equation for $\phi$ and $\mathcal{B}_{a},\,\mathcal{B}_{ab},$ and $\mathcal{B}$ are boundary terms. Therefore,
demanding invariance of $\mathcal{L}_{\mathrm{M}}$ under the infinitesimal conformal transformations (\ref{ee1})-(\ref{ee3}) we obtain
\begin{align}
\varepsilon\,e^{d}&\wedge \ast \mathcal{T}_{d}+\lambda \ast \sigma_{ab}\wedge e^{a}%
\mathrm{I}^{b}\mathrm{d}\varepsilon-\alpha\varepsilon \Phi\phi\\&
+\mathrm{d}\left(
\mathcal{B}_{a}^{\left(  2\right)  }\wedge \varepsilon e^{a}+\lambda
\mathcal{B}_{ab}^{\left(  2\right)  }\wedge e^{a}\mathrm{I}^{b}\mathrm{d}%
\varepsilon-\alpha\varepsilon \mathcal{B}^{\left(  3\right)  }\phi \right)  =0\,. \nonumber
\end{align}
It is straightforward to prove the identity%
\begin{equation}
\ast \sigma_{ab}\wedge e^{a}\mathrm{I}^{b}\mathrm{d}\varepsilon=-\mathrm{d}\left[
\varepsilon \mathrm{I}_{a}\sigma^{a}{}_{b}\ast e^{b}\right]  +\varepsilon
\mathrm{d}\left(  \mathrm{I}_{a}\sigma^{a}{}_{b}\ast e^{b}\right)  ,
\end{equation}
and therefore to conclude that%
\begin{equation}
\int_{M_{4}}\mathcal{H}^{(4)}+
\int_{\partial M_{4}}\mathcal{U}^{(3)}=0\,.
\end{equation}
where 
\begin{equation}
\mathcal{H}^{(4)}=\varepsilon \left[  e^{d}\wedge \ast \mathcal{T}_{d}+\lambda \mathrm{d}\left(
\mathrm{I}_{a}\sigma^{a}{}_{b}\ast e^{b}\right)  -\alpha\Phi\phi \right]
\end{equation}
and
\begin{equation}
\mathcal{U}^{(3)}\!\!=\! \mathcal{B}_{a}^{\left(  2\right)  }%
\wedge \varepsilon e^{a}+\lambda \!\left[  \mathcal{B}_{ab}^{\left(  2\right)
}\!\!\wedge e^{a}\mathrm{I}^{b}\mathrm{d}\varepsilon-\varepsilon \mathrm{I}_{a}\sigma
^{a}{}_{b}\ast e^{b}\right]  -\alpha\varepsilon \mathcal{B}^{\left(  3\right)  }%
\phi 
\end{equation}
Since $\varepsilon$ is arbitrary, the integrals over the bulk and over the
boundary must vanish independently. Even more, on the bulk we must have%
\begin{equation}
e^{d}\wedge \ast \mathcal{T}_{d}+\lambda \mathrm{d}\left(  \mathrm{I}_{a}\sigma^{a}%
{}_{b}\ast e^{b}\right)  -\alpha\Phi\phi=0\,.
\end{equation}
Since the trace of the stress-energy tensor corresponds to $\mathcal{T}=-\ast \left(
e^{d}\wedge \ast \mathcal{T}_{d}\right)  $, we have that%
\begin{equation}
\mathcal{T}-\lambda \ast \mathrm{d}\left(  \mathrm{I}_{a}\sigma^{a}{}_{b}\ast
e^{b}\right)  +\alpha\ast \Phi\phi=0\,.\label{Eq_Trace_Proto}%
\end{equation}

In a theory as ECSK or its current mimetic version, we have the on-shell
relationships%
\begin{align}
& \epsilon_{abcd}T^{c}\wedge e^{d}  =\kappa_{\mathrm{4}}\ast \sigma_{ab}\,,\\
&\Phi   =0\,,
\end{align}
and replacing them in Eq.~(\ref{Eq_Trace_Proto}), it lead us to the trace
value%
\begin{equation}
\mathcal{T}=\frac{2\lambda}{\kappa_{4}}\mathrm{d}^{\dag}\mathrm{I}_{a}T^{a}\,,
\end{equation}
with the four-dimensional coderivative operator $\mathrm{d}^{\dag}$ given by
$\mathrm{d}^{\dag}=\ast \mathrm{d}\ast$.
This way, we can see that the stress-energy tensor for a dilatation invariant
Lagrangian is not always traceless. It is traceless only when the torsion
vanishes or when the dilatation invariance is associated to the case
$\lambda=0$, i.e. when the spin connection remains untouched by the transformation.

\section{Summary \& comments}\label{sec8}

In summary, we have developed the closest version of mimetic gravity in first-order formalism, i.e., the mimetic version of ECKS gravity theory. Dynamics in this theory is described by the following equations of motion
\begin{align}
\mathcal{E}_{d}-\kappa_{\mathrm{4}}\ast \mathcal{\bar{T}}_{d}  &  =0\,,\\
\mathrm{d}\ast \left[  \mathcal{E}^{p}{}_{p}\mathrm{d}\phi \right]   &  =0\,,\label{coneq}\\
\mathcal{W}_{ab}  &  =0\, ,
\end{align}
where $\mathcal{E}_{a}$, $\mathcal{W}_{ab}$, and $\mathcal{\bar{T}}_{d} $ are given in eqs.~(\ref{mm10}), (\ref{mm11}), and (\ref{memt}) respectively. By construction, the system obeys the following conditions:
\begin{align}
  \mathrm{\mathring{D}}\ast \mathcal{\bar{T}}_{d}  &=0\,,\\
Z^{2}    &=1\,.\label{mce}
\end{align}
These equations reduce to the standard mimetic gravity equation when torsion $T^{a}$ vanishes. 

We have considered different possibilities of how torsion is affected by conformal transformation (\ref{m0}), all of them mapped by the parameter $\lambda$. The torsionless part of the spin connection $\mathring{\omega}^{ab}$ has a definite conformal transformation (\ref{mc1}), obtained from purely metric properties. However, we have the freedom to choose the contorsion $\kappa^{ab}$ transformation. The possibilities~(\ref{Eq_Gen_Conf_e})-(\ref{Eq_Gen_Conf_w}) range from a non-changing contorsion, to a contorsion that changes in such a way that it leaves the full spin connection invariant. Regardless of the type of transformation under consideration, dynamics enforce torsion to remain a non-propagating field, as in standard ECSK.

%%%%%%%%%%%%%%%%%%%%%%%%%%%%%%%%%%%%%%
This model may have non-trivial consequences in cosmology~\footnote{By the way, almost simultaneous with this article, another interesting cosmological application of mimetic theories of gravity involving non-vanishing torsion has been presented in \cite{Sur:2020lzd}}. Let us consider an explicit solution of equation (\ref{coneq}). For the sake of simplicity, during this section we take $c=1$ and $\Lambda=0$. Additionally, it is convenient to work in synchronous coordinates where the metric adopts the form 
 \begin{equation}
 \mathrm{d}s^{2}=-\mathrm{d}\tau^{2}+\gamma_{ij}\mathrm{d}x^{i}\mathrm{d}x^{j}\,,%
 \end{equation}
with $\gamma_{ij}$ the spatial section of the metric $g_{\mu\nu}$ (See~\cite[Chap.11]{Landau:1982dva}). As discussed in~\cite{Chamseddine:2013kea}, we take the scalar field to be the same as the hypersurfaces of constant time, namely
 \begin{equation}
 \phi\left(  x^{\mu}\right)  =\tau\,,
 \end{equation}
which naturally satisfies (\ref{mce}). In this coordinates, Eq.(\ref{coneq}) reads
 \begin{equation}
 \partial_{0}\left(  \sqrt{\gamma}\left(  G-\mathcal{T}\right)  \right)  =0\,,\label{fmet}
 \end{equation}
 and consequently
 \begin{equation}
 G-\mathcal{T}=\frac{\mathcal{C}\left(  x^{i}\right)  }{\sqrt{\gamma}}\,.
 \end{equation}
Here $\mathcal{C}$ is an integration $\tau$-constant, depending only on the spatial coordinates $x^{i}$. For a flat Friedmann Universe, the metric $\gamma_{ij}$ corresponds to
 \begin{equation}
 \gamma_{ij}=a^{2}\left(  \tau\right)  \delta_{ij}\,,\label{fmf}%
 \end{equation}
and therefore (\ref{fmet}) leads to
 \begin{equation}
 G-\mathcal{T}=\frac{\mathcal{C}\left(  x^{i}\right)  }{a^{3}\left(  \tau\right)  }\,.\label{dark}
 \end{equation}
 %%%%%%%%%%%%%%%%%%%%%%%%%%%%%%%%%%%%%
 Therefore, the scalar field coming from the conformal degree of freedom mimics a dark matter source. However, torsion is also present, and it behaves as an additional dark matter source in $G=-R$. Let us split the Ricci scalar as
  \begin{align}
 &R   =\mathring{R}+R\left(  \kappa\right)\,, \\
 &R\left(  \kappa\right)     =2\mathring{\nabla}_{\mu}\kappa^{\mu\nu}{}_{\nu}+\kappa^{\mu\gamma}{}_{ \mu}\kappa_{\gamma\nu}{}^{\nu}-\kappa^{\mu
 }{}_{\gamma\nu}\kappa^{\gamma\nu}{}_{\mu}\,,\label{rcap}%
 \end{align}
 where $\kappa^{\mu\nu}{}_{\lambda}$ is the contortion, $\mathring{R}$ is the torsionless Ricci scalar and
 $\mathring{\nabla}$ is the covariant derivative
 associated with the Christoffel symbol. Thus, using $\mathring{G}=-\mathring{R}$, we get%
 \begin{equation}
 G=\mathring{G}-R\left(  \kappa\right)  \,.\label{dark1}
 \end{equation}
 In order to evaluate Eq.(\ref{rcap}) explicitly, let us consider a spin tensor distribution
 $\sigma_{ab}$ which may be relevant at cosmological scales. Such spin tensor distribution has been considered, for instance, in~\cite{Izaurieta:2020xpk} where the Ansatz for the torsion tensor reads 
 \begin{equation}
 T_{\lambda\mu\nu}=\left[  X(\tau) (  g_{\lambda\nu}%
 g_{\mu\rho}-g_{\lambda\mu}g_{\nu\rho})  -2\sqrt{g}Y(  \tau)
 \epsilon_{\lambda\mu\nu\rho}\right]  u^{\rho}\,,\label{toran}%
 \end{equation}
 where $u^{\rho}$ is the co-moving four-velocity which in synchronous
 coordinates, and $X$ and $Y$ are arbitrary
 functions of time. These configurations can arise from a dark matter candidates, such as dark/Elko spinors, see for instance \cite{Rogerio:2016grn,Pereira:2014pqa,Boehmer:2008ah,Ahluwalia:2019etz,Ahluwalia:2016jwz,Ahluwalia:2016rwl,Ahluwalia:2015vea}. The non-vanishing components of (\ref{toran}) are
 \begin{align}
 T_{ij0} &  =X \gamma_{ij}u^{0}=a^{2}
 X  \delta_{ij}\,,\\
 T_{kij} &  =-2\sqrt{\gamma}\,Y  \epsilon_{kij}u^{0}%
 =-2a^{3} Y  \epsilon_{kij}\, .
 \end{align}
 Since
 \begin{equation}
 \kappa_{\mu\nu\lambda}=\frac{1}{2}\left(  T_{\nu\mu\lambda}-T_{\mu\nu\lambda
 }+T_{\lambda\mu\nu}\right)\,, \label{lm6}%
 \end{equation}
 we can evaluate Eq.(\ref{rcap})
 %\begin{equation}
% R\left(  \kappa\right)  =6\left(  \dot{X}+\frac{3}{2}\dot{\left(  a^{2}\right)}X+\frac{1}{2}X^{2}+a^{6}Y^{2}\right)
 %\end{equation}
 and Eqs.(\ref{dark})-(\ref{dark1}) to arrive to
 \begin{equation}
 \mathring{G}-\mathcal{T}=\frac{\mathcal{C}}{a^{3}}+3!\left[  X\left(  \mathcal{S}%
 +3\mathcal{H}\right)  +X^{2}+Y^{2}\right]\,,
 \end{equation}
 where $\mathcal{C}=\mathcal{C}\left(  x^{i}\right)  $, $\mathcal{S}\left(
 \tau\right)  =\frac{\dot{X}}{X}$ and $\mathcal{H}\left(  \tau\right)
 =\frac{\dot{a}}{a}\,$. From here the role of the spin-tensor as dark matter becomes evident.
 
  In the mimetic ECSK theory, there are three dark matter species. The first is the dark matter by itself, with a non-vanishing spin tensor. The second arises from isolating the conformal mode in a covariant way, and the last comes from the torsional degrees of freedom. 
  %The construction of mimetic ECKS gravity compatible with cosmological principles is currently a work in course and will be presented somewhere else. 

\section{Acknowledgments}
We are grateful to
Cristian Martínez,
Cristóbal Corral,
Subhasish Chakrabarty,
Amitabha Lahiri,
Alfredo Pérez,
and
Adolfo Cisterna
for many enlightening conversations and references.  P. M. is supported by Agencia Nacional de Investigación (ANID) through the grant 21161574. OV acknowledges VRIIP UNAP for financial support through Project VRIIP0062-19.
 FI ad PS acknowledges financial support from the Chilean government through Fondecyt grant 1180681. NM, acknowledges Fondecyt grant 11180894.
%%%%%%%%%%%%%%%%%%%%%%%%%%%%%%%%%%%%%%%%%%%%%%%%%%%%%%%%%%%%%%%%%%%%%%%%%%%%%%%%

%\bibliographystyle{utphys}
%\bibliographystyle{plain}
\bibliography{GB2019}

%%%%%%%%%%%%%%%%%%%%%%%%%%%%%%%%%%%%%%%%%%%%%%%%%%%%%%%%%%%%%%%%%%%%%%%%%%%%%%%%

\end{document}